\documentclass[aip,jap,reprint,amsmath]{revtex4-1}

\usepackage{graphicx}
\usepackage{dcolumn}
\usepackage{bm}
\usepackage{mathrsfs}

\begin{document}

\title{A computationally efficient method for calculating the maximum conductance of disordered  networks: Application to 1-dimensional conductors.}

\author{Luiz F. C. Pereira}
\email{pereirlf@tcd.ie}
\affiliation{School of Physics and CRANN, Trinity College Dublin, Dublin 2, Ireland}
\author{C. G. Rocha}
\affiliation{Institute for Materials Science and Max Bergmann Center of Biomaterials, Dresden University of Technology, D-01062 Dresden, Germany}
\author{A. Lat\'e}
\affiliation{Instituto de F\'{\i}sica, Universidade Federal Fluminense, Niter\'oi, Brazil}
\author{M. S. Ferreira}
\email{ferreirm@tcd.ie}
\affiliation{School of Physics and CRANN, Trinity College Dublin, Dublin 2, Ireland}

\date{\today}

\begin{abstract}
Random networks of carbon nanotubes and metallic nanowires have shown to be very useful in the production of transparent, conducting films. The electronic transport on the film depends considerably on the network properties, and on the inter-wire coupling. Here we present a simple, computationally efficient method for the calculation of conductance on random nanostructured networks. The method is implemented on metallic nanowire networks, which are described within a single-orbital tight binding Hamiltonian, and the conductance is calculated with the Kubo formula. We show how the network conductance depends on the average number of connections per wire, and on the number of wires connected to the electrodes. We also show the effect of the inter-/intra-wire hopping ratio on the conductance through the network. Furthermore, we argue that this type of calculation is easily extendable to account for the upper conductivity of realistic films spanned by tunneling networks. When compared to experimental measurements, this quantity provides a clear indication of how much room is available for improving the film conductivity. 
\end{abstract}

\maketitle

\section{Introduction}
\begin{figure}
\includegraphics[width = 5cm] {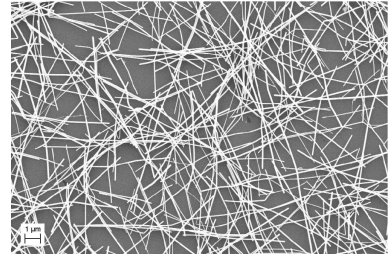}
\caption{SEM image of a silver nanowire film, showing the disordered complex network. Courtesy of J. N. Coleman.}
\label{fig:film}
\end{figure}

The current research on flexible, transparent and conductive thin films is driven by their huge potential as transparent electrodes, with many promising applications in the construction of low-power foldable electronic displays \cite{nomura,lewis,jackson,Wu:04,Sukanta:09}.  Metallic nanowires (NW) are the most prominent materials of choice to promote electronic transport in these films. In this case, electrons move between several NW forming a disordered tunneling network which spans the entire film. Fig. \ref{fig:film} displays a microscopy image of one such film. The conductivity is primarily limited by the junction resistance that exists between neighboring NW. In fact, this has been confirmed by several studies reporting enhancements in the film conductivity as a result of improvements in the inter-wire contacts \cite{Geng:07, Boland:09}. However, in addition to the inter-wire resistance, the network morphology also plays a role in limiting the film conductivity. This means that no matter how much progress is made in lowering the junction resistance, there should be a maximum value for the conductivity, which is regulated by the network topology. 

From a theoretical perspective, the description of electronic transport across these films involves the flow of charge through a percolating network made of finite-sized 1-dimensional (1D) conductors connected to source and drain electrodes. Computational studies of this nature have so far concentrated on analyzing this phenomenon on two very different length scales. The first of these approaches is based on the ohmic-behavior displayed by percolating networks of finite-length conducting rods. In this case the rods are represented by length-dependent ohmic resistances, connected to each other by inter-wire resistors, and the overall network resistivity is obtained by solving Poisson's equation \cite{Ural:07, Ural:09}. A somewhat similar method has also been used to describe the film conductance when the inter-wire resistance is fully dominant. In this case, common when the networks are made of carbon nanotubes (CNT) for instance, it is possible to solve Kirchhoff's equations to obtain the potential drop across the network \cite{Lyons:08}. Both methods agree qualitatively well with experimental measurements. 

It is also possible to approach the problem from an atomistic viewpoint, most notably by means of {\it ab-initio} calculations coupled with non-equilibrium Green function methods. The computational cost of these calculations is considerably high, and the absence of periodicity on the systems severely limits the studies performed with this technique. Nonetheless, recent studies have provided results for the intra- and inter-wire conductance in the case of a single carbon nanotube junction in the presence of O$_2$ and N$_2$ molecules \cite{Mowbray:09}. The use of such a method for obtaining the quantum conductance between individual wires is able to shed some light on the quality of the junction resistances but it is unable to reproduce the general observed features since it does not include any disorder effects. 

The alternative of carrying out a fully atomistic transport calculation within a heavily disordered environment is very desirable but it is currently too computationally demanding and calls for a compromise if one wishes to combine the two features. One such a way consists of using semi-empirical methods for the electronic structure of the wires,  {\it e.g.} tight-binding models, whose parameters are obtained from {\it ab-initio} calculations \cite{charlier04, cgrocha}. Because this method is significantly less demanding, it is possible to include disorder effects which take into account extensive configurational averaging procedures. This combination has been used recently to establish the upper value for the conductivity of CNT-based films with the intention of guiding the experimental optimization of these films \cite{Pereira:09}. By comparing the experimentally measured conductivities with the corresponding upper value, one knows how much room there is for improvement, usually attempted by chemical treatments that can reduce the inter-wire junction resistance.  This valuable piece of information has been used, for instance, to establish that the best measured values of conductivities for nanotube-based films are too close to the upper limit predicted by this technique, which suggests that further improvements in their transport properties are unlikely.  

The goal of the present work is therefore to present a simple way in which the upper value for the conductance of heavily disordered networks can be obtained. Since we are interested in the best-case scenario in which the electronic transmission is at its maximum, we must eliminate potential sources of scattering and decoherence such as structural imperfections, impurities, and interaction with other quasiparticles. In this situation, it is appropriate to consider a purely ballistic regime of transport within the network, which calls for a quantum description of the conductance. Furthermore, our method is supposed to be easily transferable to account for a variety of different materials including nanotubes, noble-metal wires and graphene \cite{Ural:09b}, to name but a few. With that in mind, we describe the transport calculations within the Kubo formalism expressed in terms of single-particle Green functions (GF). The advantage of using this method is that once the ``wires'' are chosen, the corresponding GF is automatically defined, making results easily transferable between different cases. Ballistic transport calculations in such an idealized environment provide the maximum possible conductance for these disordered systems and are easily translated into an upper limit for the conductivity of films made of similar networks. In particular, these limits can be studied in terms of film porosity and in terms of the aspect ratio of the components of the film. 

\section{Model}
\label{sec:model}

We start by defining the system we are considering and how we intend to describe its electronic structure. We assume a collection of nanosized elements that are disorderly oriented and distributed forming a complex network. This structure is modeled by a random graph \cite{Erdos:59,Bollobas-book}, which is defined by its total number of nodes ($N$) and the average number of connections per node ($\langle \alpha \rangle$). The graph is generated such that each node has its own number of connections, while the distribution of connections obeys a Poisson distribution. Once the interconnected network has been generated, a number of nodes ($\langle \alpha_E \rangle$) is chosen to connect the network to each of the source and drain electrodes. Notice that in order to avoid short-circuits, all nodes in the network are explicitly prevented from being in contact with both electrodes at any given configuration.

The electronic structure of individual constituents of the system can be described by a tight-binding Hamiltonian of the form
\begin{equation}
{\mathcal H_\beta} = \sum_{i} |\beta, i \rangle \epsilon_0 \langle \beta, i | + \sum_{ \langle i,j \rangle } |\beta, i \rangle \gamma_0 \langle \beta, j |,
\label{TB-H}
\end{equation}
where $|\beta, i \rangle$ represents an electron on atom $i$ of an element labeled by $\beta$, $\epsilon_0$ is the on-site energy, $\gamma_0$ is the hopping integral, and the sum over $\langle i,j \rangle$ is over nearest-neighboring atoms. This general form can describe several types of nanostructures, depending on the choice of atomic structure and on the orbital degrees of freedom represented by the states $| i \rangle$. Moreover, it is straightforward to generalize this Hamiltonian to the case when there is a collection of ${N}$ such nanostructures through a simple sum ${\mathcal H}_0 = \sum_{\beta=1}^{N} {\mathcal H_\beta}$ as long as there is no interaction between elements, {\it i.e.}, assuming they are not interconnected. It is simple to connect the nanostructures if we introduce a potential bridging any two that are in contact. This inter-element coupling appears in the form of an electronic hopping term $\gamma^\prime$ in the Hamiltonian given by
\begin{equation}
 V_{\beta,\beta^\prime} = | \beta, j \rangle \gamma^\prime \langle \beta^\prime , j^\prime |,
\end{equation}
where $\beta$ and $\beta^\prime$ label the connecting nanostructures, and $j$ and $j^\prime$ identify the atoms in each nanostructure. Notice that $\beta$ and $\beta^\prime$ must be different while $j$ and $j^\prime$ are allowed to have the same value.

Associated with the Hamiltonian ${\mathcal H}_\beta$, there are the retarded($+$) and advanced($-$) single-particle GF, defined as
\begin{equation}
{\mathcal G}^{\pm} = [E^\pm - {\mathcal H_\beta}]^{-1},
\label{eq:GF}
\end{equation}
where $E^\pm = E \pm \imath \eta$, $\eta$ being a small positive imaginary part added to the electron energy $E$. It is worth emphasizing that the choice of Hamiltonian automatically defines the corresponding GF. So, for the sake of simplicity, in what follows we present results for wires consisting of monatomic chains with single orbital atoms. Wires of this type are known to carry a single quantum of conductance ( $\Gamma_0 = 2e^2/h$) but it is very simple to introduce orbital degeneracy to describe wires capable of carrying more conducting channels.

In the case of finite-sized 1D atomic chains, it is possible to derive an analytic expression for its GF \cite{Economou-book}. The $(i,j)$-th matrix element of the GF of a finite monatomic chain with $L$ atoms described by the Hamiltonian in Eq.(\ref{TB-H}) is given by 
\begin{equation}
{\mathcal G}^{\pm}_{ij} (E) = \frac{2}{L} \sum_{m=1}^{L} \frac{\sin(i\frac{m \pi}{L+1}) \sin(j\frac{m \pi}{L+1})}{E^\pm  - \left[\epsilon_0 + 2 \gamma_0 \cos(\frac{m \pi}{L+1})\right]},
\label{1D-GF}
\end{equation}
which is very general and can be used to calculate any matrix elements of the GF, as needed. Notice that the expression above recovers the familiar result of infinite-chain density of states (DOS) in the limit $L \rightarrow \infty$. For reasons that will become clear later, we also present an analytical expression for the surface matrix element of the GF for a semi-infinite monatomic chain, which is written as
\begin{equation}
\label{eq:surfaceGF}
 \mathcal S^\pm (E) = \frac{(E^\pm - \epsilon_0) \pm \left[(E^\pm - \epsilon_0)^2-4 |\gamma_0|^2\right]^{1/2} }{2 |\gamma_0|^2},
\end{equation}
where for the $\pm$ sign, one has to choose the sign which provides a positive DOS. 

Note that Eq.(\ref{1D-GF}) describes the GF of a single isolated wire. The GF of an ensemble of ${N}$ disconnected wires is easily represented by a matrix whose diagonal blocks are given by the GF defined in Eq.(\ref{eq:GF}). In other words, the GF of the disconnected network $g^\pm$ is written as 
\begin{eqnarray}
g^\pm =  \left(
\begin{array}{ccccc}
{\mathcal G}^\pm & 0 & 0 & 0 & 0 \\
0 & {\mathcal G}^\pm  & 0 & 0 & 0 \\
0 & 0 & {\mathcal G}^\pm & 0 & 0 \\
0 & 0 & 0 & {\mathcal G}^\pm & 0 \\
0 & 0 & 0 & 0 & {\mathcal G}^\pm \\
\end{array}
\right),
\label{matrix-g}
\end{eqnarray}
where each element in the matrix above is actually a matrix itself whose dimension is defined by the length of the finite chain representing the NW. The 5-block matrix appearing in Eq.(\ref{matrix-g}) is written for illustration purposes only. In general, it is a much larger matrix whose number of blocks equals the total number of wires ${N}$. The effect of the interaction between wires is easily computed using Dyson's equation, which can be written as
\begin{equation}
G^\pm = g^\pm + g^\pm V  G^\pm,
\label{eq:dyson}
\end{equation}
where $G^\pm$ is the network GF, $g^\pm$ is the GF of the disconnected wires defined in Eq.(\ref{matrix-g}) and $V$ is the inter-wire coupling matrix. In this particular illustrative example the matrix $V$ may look something like
\begin{eqnarray}
V =  \left(
\begin{array}{ccccc}
0 & 0 & 0 & 0 & V_{1,5} \\
0 & 0 & 0 & V_{2,4} & V_{2,5} \\
0 & 0 & 0 & 0 & 0 \\
0 & V_{4,2} & 0 & 0 & 0 \\
V_{5,1} & V_{5,2} & 0 & 0 & 0 \\
\end{array}
\right),
\label{matrix-V}
\end{eqnarray}
indicating that wire 1 has a single connection to wire 5; wire 2 is connected to wires 4 and 5;  wire 4 is linked only with wire 2 and wire 5 is connected with wires 1 and 2. Note that in this pictorial representation wire 3 is not connected to any of its counterparts, a fact that fills the third row and third column of the matrix $V$ with zero block matrices. In practical terms, the corresponding row/column in the matrix $G^\pm$ will not involve any $V$-dependent terms and is simply given by ${\mathcal G}^\pm$. It is easy to conclude that similar behavior arises in terms of the intra-wire indices, that is, a connection between atom $j$ of wire $\beta$ and atom $j^\prime$ of wire $\beta^\prime$ will only involve the corresponding matrix elements of the $g$-matrix. Therefore, if a given NW is connected to three neighbors at intra-wire sites $i$, $j$ and $k$, the matrix elements ${\mathcal G}^\pm_{i,i}$, ${\mathcal G}^\pm_{j,j}$,  ${\mathcal G}^\pm_{k,k}$, ${\mathcal G}^\pm_{i,j,}$, ${\mathcal G}^\pm_{i,k}$, ${\mathcal G}^\pm_{j,k}$ (and the respective adjoint elements) are the only matrix elements appearing in the expression for the network GF of Eq.(\ref{eq:dyson}). This can be an enormous simplification when compared with the full diagonalization method. Rather than inverting enormous matrices of the order of ${N} \times L$, this technique allows us to express the network GF in terms of inverse operations of much smaller matrices, and whose sizes are defined by the number of connections in the system, {\it i.e.}, matrices of order $N \times \langle \alpha \rangle$. This method becomes extremely advantageous when the Kubo formula for the conductance is used, as we show in the following section.

\section{Kubo Formalism}
\label{sec:conductance}

\begin{figure}
\includegraphics[width = 6cm] {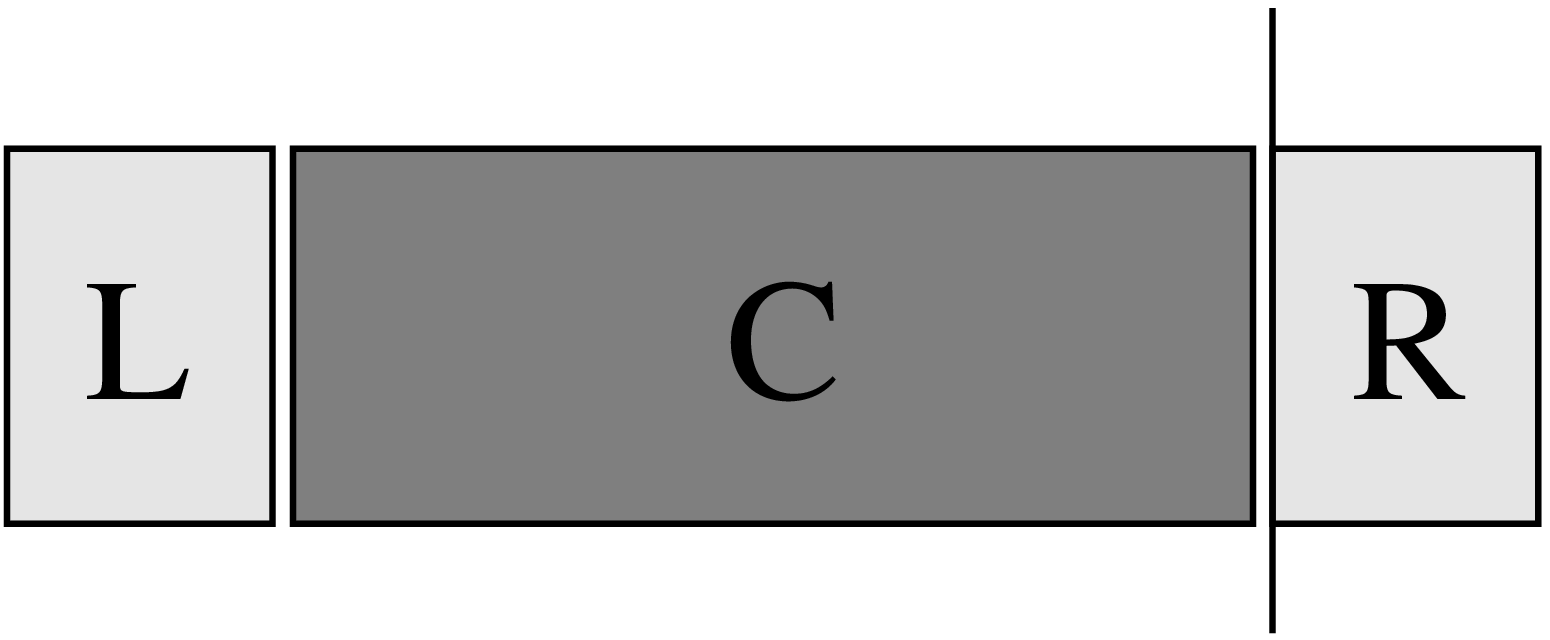} \\
\includegraphics[width = 6cm] {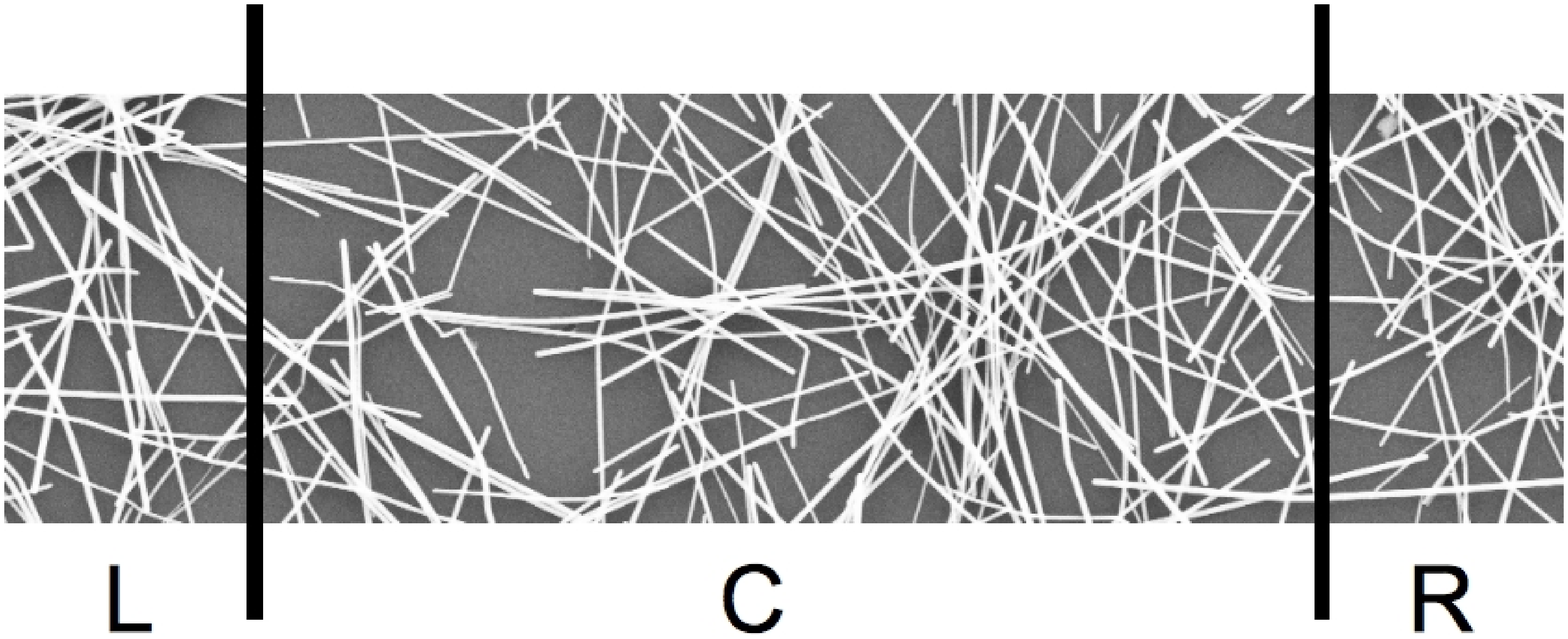}
\caption{Schematic representation of the system. The scattering region C contains the disordered array of finite nanowires. Each perfect lead L and R contain several independent semi-infinite wires. On the upper panel the vertical line indicates the location of the cleavage plane through which the conductance is calculated.}
\label{fig:system}
\end{figure}

The conductance of the system can be calculated by the Kubo formalism, which is equivalent to the more familiar Landauer method \cite{datta-book}. The Kubo formalism provides a simple expression for the conductance by obtaining the net current that travels across a reference plane in the system, often referred to as the cleavage plane, placed between two parts of the system. When this current is expressed in terms of a few GF matrix elements of the system, the calculation becomes extremely simple. In this formalism the system under study can be divided into three parts: a central scattering region (C) sandwiched by two perfect semi-infinite leads (L,R), as illustrated schematically in the top panel of Fig. \ref{fig:system}. The location of the cleavage plane, represented by the vertical line, is arbitrary and can be chosen in the most convenient location for the case in question. In our setup the conductance of the system can be calculated by placing the cleavage plane between either one of the leads, and the central region. The leads are represented by semi-infinite objects, since it is necessary to have particle reservoirs where the charge can move to and from. When expressed in terms of GF, this method depends only on the GF matrix elements located on both sides of the cleavage plane, as if the two halves were totally disconnected. 

The implementation presented here is a generalization of the usual setup described above, in which the L, C and R regions are actually made of subsets of the entire network. The bottom panel of Fig. \ref{fig:system} shows how the system can be separated into leads and a scattering region. Analogously to the standard case in which charge reservoirs are needed, here we can impose that every wire touching the electrode regions L and R is replaced by a semi-infinite wire, which provides source and drain for the moving charges. The semi-infinite wires in the leads do not interact with each other, and are allowed to connect to only one finite conductor in the network. Restricting semi-infinite conductors to only one connection to the network is not a fundamental requirement, but it is a safe and direct way of avoiding short-circuits which could provide spuriously high values for the conductance. It is also important to bear in mind that the finite wires connected to the semi-infinite ones can have numerous connections themselves. Translating Fig. \ref{fig:system} into words, regions L and R contain $\langle \alpha_E \rangle$ semi-infinite wires, and region C contains a disordered array of $N$ finite wires, where each one has a random number of connections, with an average number of connections per wire given by $\langle \alpha \rangle$. 

Having placed the cleavage plane between the regions C and R, the Kubo formula for the zero-bias conductance is written as \cite{Mathon:97}
\begin{equation}
\label{eq:kubo}
\begin{split}
\Gamma(E) & = \left( \frac{4 e^2}{h} \right) \times \\
& \Re e \left\{ \mathrm{Tr}
\left[ \tilde G_{C}  V_{CR} \tilde G_{R}  V_{CR}^\dagger
 - V_{CR} \tilde G_{RC} V_{CR} \tilde G_{RC} \right] \right\},
\end{split}
\end{equation}
where all the GF are calculated at the Fermi energy $E_F$, and $\tilde G(E)$ are the causal GF, defined as
\begin{equation}
\tilde G(E) = \frac{1}{2 \imath} \left[ G^-(E) - G^+(E) \right].
\end{equation}

It is worth stressing once again the advantages of using the set of Eqs. (\ref{1D-GF}) to (\ref{matrix-V}) to derive the GF of the entire network. As such the matrix elements required to evaluate the conductance in the equation above are easily calculated. In particular, we emphasize the massive reduction introduced in the GF matrix to obtain only the relevant elements. Each finite wire can be represented by a GF matrix which contains only the elements corresponding to atoms interacting with other wires (or the leads) and the propagators between those atoms. For example, a wire with $L$ atoms is described by a GF matrix with $L^2$ elements, but if this wire interacts with only $\alpha$ other wires in the network, then its GF matrix can be reduced to just $\alpha^2$ elements. The same reduction is applied to the GF of the leads. Each semi-infinite conductor in the leads is represented by a single matrix element, corresponding to the atom which connects to the array. Furthermore, with the use of simple analytic expressions for the GF matrix elements, the procedure to calculate the network propagators is tremendously simplified. Moreover, because our expressions are written in terms of GF, it is straightforward to replace them with GF associated with other types of nanostructures. For instance, recent calculations have been carried out with numerically evaluated GF for carbon nanotubes \cite{Pereira:09}, and could possibly be also done for graphene \cite{Ural:09b}. Finally, another advantage worth mentioning is that the simplicity with which we treat the system allows us to evaluate results for several samples in parallel. This is paramount in calculations performed in disordered environments since several samples are required to obtain statistically significant results. For each data point presented in the next section, several independent configurations were generated, and all the error bars (calculated from statistical errors in the samples) are smaller than symbol sizes.

\section{Results and Discussion}
\label{sec:results}

In order to better understand the role of each parameter on the electronic transport across disordered networks, we study the effects of $\langle \alpha \rangle$, $\langle \alpha_E \rangle$, and inter-wire hopping, independently.

We begin by analyzing the effect of the connectivity between wires on the conductance of the network. If the number of connections per node is too small,{\it i.e.} $1 \le \langle \alpha \rangle < 2$, isolated islands tend to form, instead of a connected percolating network, and thus the conductance of the system is expected to vanish on the thermodynamic limit ($N \rightarrow \infty$). On the other hand, for $\langle \alpha \rangle = 2$, $\approx 80\%$ of the wires are interconnected, and a percolating cluster is always formed for $\langle \alpha \rangle \ge 2$. However, the electrodes must be connected to this largest cluster in order to form a
conducting system. In particular, for $\langle \alpha \rangle = 2$, there is a $20\%$ chance that any electrode wire will be connected to one of the smaller isolated islands, and will thus not contribute to the electronic transport across the network. For comparison, we state that for $\langle \alpha \rangle = 4$ the fraction of interconnected nodes is $98\%$, whereas for $\langle \alpha \rangle = 6$ this fraction becomes $99.5\%$.

\begin{figure}
\includegraphics[width = 7cm] {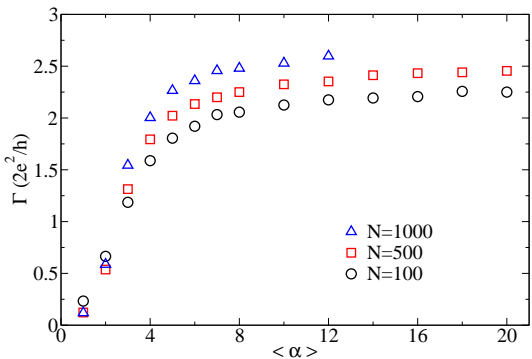}
\caption{(Color online) Conductance at a fixed energy as a function of the connectivity of the network. Electronic transmission increases as more conducting paths become available. Legends indicate total number of wires. The length of the wires is fixed at $L=1000$, and there are $\langle \alpha_E \rangle = 6$ connections to each electrode. Data averaged over $1000$ independent configurations.}
\label{fig:cond_vs_alpha}
\end{figure}

In Fig. \ref{fig:cond_vs_alpha} we show the dependence of the conductance at a fixed electronic energy as a function of the connectivity of the network, for networks with three different number of components. The data was obtained considering wires of fixed length $L=1000$ atoms, and the number of wires connected to each electrode (semi-infinite wires) was fixed at $\langle \alpha_E \rangle = 6$. The curves emphasize the influence of the network connectivity on the conductance, as the former is increased from its lower limit $\langle \alpha \rangle=1$, up to values well beyond the percolation limit. Each data point is an average over $1000$ independent configurations, and all realizations were considered in the averaging procedure. The increase in conductance is a consequence of the increase in the number of available conducting paths for the electronic transport across the network. However, as more of these conducting paths become available, the probability of electrons returning to their original electrode also increases (zero-bias calculation), which brings the system to a steady state where the conductance of the network saturates. Furthermore, the increment in conductance with the number of wires $N$ seen on Fig. \ref{fig:cond_vs_alpha}, is also caused by the expansion of conducting paths, since the total number of unique connections in the network is given by $\langle \alpha \rangle N / 2$.

Another major factor on the electronic transport of random nanowire networks is the number of wires connected to the electrodes. It is important to remember that each wire connected to an electrode provides a charge reservoir from where electrons come, and to where they finally go. Moreover, each one of these electrode wires supports only one quantum of conductance, which introduces a bottleneck effect on the system. For example, if $\langle \alpha_E \rangle = 1$, no matter how many conducting paths are present on the network, only one conduction channel is available for electrons moving from the network towards the drain electrode, and the maximum conductance is therefore limited to one quantum of conductance $\Gamma_0 = 2e^2/h$. On the other hand, for any connectivity beyond the percolation limit several conduction paths will be available on the network, thus increasing $\langle \alpha_E \rangle$ will certainly improve the conductance of the network. 

\begin{figure}
\includegraphics[width = 7cm]{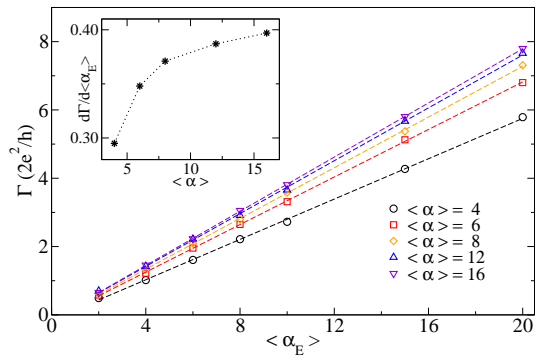}
\caption{(Color online) Network conductance as a function of the number of connections to electrodes, $\langle \alpha_E \rangle$. Inset: slope of the linear-fitted lines increases with the network connectivity $\langle \alpha \rangle$, up to a saturation point. Data is for $N=100$, and $L=1000$, averaged over at least $200$ independent configurations.}
\label{fig:cond_vs_alphaE}
\end{figure}

On Fig. \ref{fig:cond_vs_alphaE} we present the direct effect of  $\langle \alpha_E \rangle$ on the network conductance, for different values of connectivity $\langle \alpha \rangle$. The conductance of the whole network depends linearly on the number of connections to semi-infinite conductors, with a slope that varies with the connectivity of the network. The dashed lines are linear least-square fits to the data. Notice that as $\langle \alpha_E \rangle$ is increased the response in conductance is quite considerable, which clearly indicates that once a percolating network is in place, the system conductance is limited only by $\langle \alpha_E \rangle$. On the inset, we show the variation of ${d \Gamma}/{d \langle \alpha_E \rangle}$ with the connectivity $\langle \alpha \rangle$, obtained from the slope of the linear-fitted lines. The slope reaches a maximum value for large enough connectivities, when an increase in the number of conducting paths available to the electrons does not alter the conductance of the network any further.

The conductance of random nanowire networks is expected to be strongly dependent on the probability of electrons to jump between neighboring wires, which is related to the inter-wire hopping parameter. This inter-wire hopping is defined in analogy to the intra-wire hopping integral $\gamma_0$, and is denoted by $\gamma^\prime$. In the tight-binding model, the hopping integral represents the strength of the bond between two atoms in a molecule or solid, therefore a higher value indicates a stronger coupling. It is well known that in metallic nanowire/carbon nanotube films, the resistance arises mostly from inter-wire electron tunneling, and thus, films with better coupled elements have lower overall resistance. 

\begin{figure}
\includegraphics[width = 7cm]{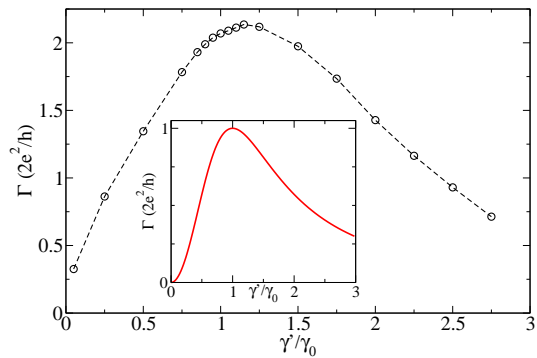}
\caption{(Color online) Network conductance as a function of inter-wire/intra-wire coupling ratio. Maximum conductance is obtained around $\gamma^\prime /\gamma_0 = 1.15$. $N=100$, $\langle \alpha \rangle = 8$, $\langle \alpha_E \rangle = 6$, $L=1000$, and $1000$ independent configurations. Inset: Conductance of two semi-infinite chains connected by a hopping $\gamma^\prime$. Further details on the text. }
\label{fig:cond_vs_gamma}
\end{figure}

The inter-atomic (intra-wire) hopping is the only parameter in the tight-binding model considered here, and so it is only natural that we choose to define the inter-wire hopping in terms of this parameter. That is to say, we analyze the influence of the ratio $\gamma^\prime/\gamma_0$ on the conductance of the network. If the ratio is too small ($\gamma^\prime /\gamma_0 << 1$) or too large ($\gamma^\prime /\gamma_0 >> 1$) the mismatch between the couplings causes an adverse effect on the electronic transport through the system, lowering the conductance from its maximum possible value. In particular, considering only two semi-infinite wires connected to each other at the surface by a hopping $\gamma^\prime$ (forming an infinite linear chain), we can see on the inset of Fig. \ref{fig:cond_vs_gamma} that the maximum conductance occurs when the inter-/intra-wire couplings are equal to each other, {\it i.e.}, $\gamma^\prime/\gamma_0 = 1$. This is not surprising if we bear in mind that any contrast in the Hamiltonian parameters is a source of additional scattering. However, on the main panel of Fig. \ref{fig:cond_vs_gamma} we show the results for network conductance as a function of  $\gamma^\prime /\gamma_0$ for networks with $N=100$, $\langle \alpha \rangle = 8$, $\langle \alpha_E \rangle = 6$, and $L=1000$. In this case, it can be observed that the maximum conductance occurs for ratios slightly larger than $1$, which results from the fact that finite-sized wires hanging loose off the conducting path are capable of causing similar scattering events \cite{Grimm:05}. Moreover, we have observed that the actual maximum value of $\Gamma$ and the ratio at which it occurs depends slightly on the fixed value chosen for the electronic energy.
 
\section{Conclusion}
\label{sec:conclusion}

In summary, we have presented a general computational procedure for the calculation of the quantum conductance in disordered arrays of nanoscale conductors, in the ballistic limit. This method was implemented within the tight-binding approximation for the electronic structure of individual 1D conductors. We have shown how the conductance of the network increases with the connectivity, up to a saturation point. The linear dependence of the conductance with the number of connections to electrodes, for a fixed connectivity, was also presented. Finally the dependence of the conductance on the inter-wire/intra-wire hopping ratio was calculated, and the optimum point identified. 

In order to calculate the maximum conductivity of films made of specific nanostructures (NW, CNT, graphene, etc), it is necessary to define a geometric model which captures the microscopic arrangement in the film. The connectivities of the film can be extrated from the model, in terms of quantities like the volume fraction (or density) of the film, and the aspect ratio of the individual components \cite{Lyons:08}. In following this recipe we have been able to calculate the upper bound for the conductivity of films spanned by an array of disordered carbon nanotubes. Most remarkably, this value appears to be pretty close to the state-of-the-art experimental measurements reported so far \cite{Pereira:09}. A similar strategy must be followed in order to calculate the upper bound for the conductivity of films made of other nanostructures. 

\acknowledgements
L.F.C.P. and M.S.F. acknowledge financial support from Science Foundation Ireland (SFI). A.L. acknowledges partial financial support from Rede Nacional de Nanotubos. L.F.C.P. would like to thank the SFI/HEA Irish Centre for High-End Computing (ICHEC) and the Trinity Centre for High Performance Computing (TCHPC) for the provision of computational facilities and support.


\thebibliography{}
\bibitem{nomura} K. Nomura, H. Ohta, A. Takagi, T. Kamiya, M. Hirano, and H. Hosono, Nature {\bf 432}, 488 (2004). 

\bibitem{lewis} J. Lewis, S. Grego, B. Chalamala, E. Vick, and D. Temple, Appl. Phys. Lett. {\bf 85}, 3450 (2004). 

\bibitem{jackson} W.B. Jackson, R.L. Hoffman, and G.S. Herman, Appl. Phys. Lett. {\bf 87}, 193503 (2005). 

\bibitem{Wu:04}  Z.C. Wu, Z.H. Chen, X. Du, J.M. Logan, J. Sippel, M. Nikolou, K. Kamaras, J.R. Reynolds, D.B. Tanner, A.F. Hebard, and A.G. Rinzler, Science {\bf 305}, 1273 (2004).

\bibitem{Sukanta:09} S. De, T.M. Higgins, P.E. Lyons, E.M. Doherty, P.N. Nirmalraj, W.J. Blau, J.J. Boland, and J.N. Coleman, ACS Nano {\bf 3}, 1767 (2009).

\bibitem{Geng:07}   H.Z. Geng, K.K. Kim, K.P. So, Y.S. Lee, Y.Chang, and Y.H. Lee, J. Am. Chem. Soc. {\bf 129}, 7758 (2007).

\bibitem{Boland:09} P.N. Nirmalraj, P.E. Lyons, S. De, J.N. Coleman and J.J. Boland, Nano Lett.  {\bf 9}, 3890 (2009).

\bibitem{Ural:07} A. Behnam and A. Ural, Phys. Rev. B {\bf 75}, 125432 (2007).

\bibitem{Ural:09} J. Hicks, A. Behnam, and A. Ural, Phys. Rev. E {\bf 79}, 012102 (2009).

\bibitem{Lyons:08} P.E. Lyons, S. De, F. Blighe, V. Nicolosi, L.F.C. Pereira, M.S. Ferreira, and J.N. Coleman, J. Appl. Phys.  {\bf 104}, 044302 (2008).

\bibitem{Mowbray:09} D.J. Mowbray, C. Morgan, and K.S. Thygesen, Phys. Rev. B. {\bf 79}, 195431 (2009).

\bibitem{charlier04} S. Latil, S. Roche, D. Mayou and J.-C. Charlier, Phys. Rev. Lett. {\bf 92}, 256805 (2004).

\bibitem{cgrocha} C.G. Rocha, A. Wall, A.R. Rocha and M.S. Ferreira, J. Phys.: Cond. Matter {\bf 19}, 346201 (2007).

\bibitem{Pereira:09} L.F.C. Pereira, C.G. Rocha, A. Latg\'e, J.N. Coleman and M.S. Ferreira, Appl. Phys. Lett. {\bf 95}, 123106 (2009).

\bibitem{Ural:09b} J. Hicks, A. Behnam, and A. Ural, Appl. Phys. Lett. {\bf 95}, 213103 (2009).

\bibitem{Erdos:59} P. Erd\"os and A. R\'enyi, Publicationes Mathematicae {\bf 6}, 290 (1959).

\bibitem{Bollobas-book} B. Bollob\'as, ``Random Graphs", Cambridge University Press (2001).

\bibitem{Economou-book} E.N. Economou, ``Green's Functions in Quantum Physics", Springer (2006).

\bibitem{datta-book} S. Datta, ``Electronic Transport in Mesoscopic Systems", Cambridge University Press (1997).

\bibitem{Mathon:97} J. Mathon, A. Umerski and M. Villeret, Phys. Rev. B. {\bf 55}, 14378 (1997).

\bibitem{Grimm:05} D. Grimm, A. Latg\'e, R.B. Muniz, and M.S. Ferreira, Phys. Rev. B. {\bf 71}, 113408 (2005).
\end{document}